\documentclass{ws-procs11x85}

\usepackage{multicol}
\usepackage{epsfig,amsmath,subfigure,latexsym,amssymb}

\def\Journal#1#2#3#4{{#1} {\bf #2}, #3 (#4)}

\def\lsi{\raise0.3ex\hbox{$<$\kern-0.75em\raise-1.1ex\hbox{$\sim$}}}
\def\gsi{\raise0.3ex\hbox{$>$\kern-0.75em\raise-1.1ex\hbox{$\sim$}}}

\newcommand{\lsim}{\mathop{\lsi}}
\newcommand{\gsim}{\mathop{\gsi}}
\newcommand{\Z}{\mathbb{Z}}

\makeindex
\begin{document}

\title{THE CONTINUUM LIMIT OF THE NON-COMMUTATIVE $\lambda \phi^{4}$ MODEL}

\author{W. BIETENHOLZ \and F. HOFHEINZ}

\address{Institut f\"{u}r Physik, Humboldt Universit\"{a}t zu Berlin,
Newtonstr. 15, D-12489 Berlin, Germany
\\E-mail: bietenho@physik.hu-berlin.de, hofheinz@physik.hu-berlin.de}

\author{J. NISHIMURA}

\address{High Energy Accelerator Research Organization (KEK),
1-1 Oho, Tsukuba 305-0801, Japan\\E-mail: jnishi@post.kek.jp}

%%%%%%%%%%%%%%%%%%%%%%%%%%%%%%%%%%%%%%%%%%%%%%%%%%%%%%%%%%%%%%%%%%%%%%%%%
% You may repeat \author \address as often as necessary             %
%%%%%%%%%%%%%%%%%%%%%%%%%%%%%%%%%%%%%%%%%%%%%%%%%%%%%%%%%%%%%%%%%%%%%%%%%

\maketitle\abstract{ We present a numerical study of the $\lambda \phi^{4}$
model in three Euclidean dimensions, where the two spatial 
coordinates are non-commutative (NC).
We first show the explicit phase diagram of this model on a lattice. 
The ordered regime splits into a phase of uniform order and a ``striped 
phase''. Then we discuss the dispersion relation, which allows us
to introduce a dimensionful lattice spacing. 
Thus we can study a double scaling limit to zero lattice spacing 
and infinite volume, which keeps
the non-commutativity parameter constant.
The dispersion relation in the disordered phase stabilizes in this
limit, which represents a non-perturbative renormalization.
From its shape we infer that
the striped phase persists in the continuum, and we observe
UV/IR mixing as a non-perturbative effect.}

\begin{multicols}{2}

\baselineskip=13.07pt
\section{The Non-Commutative Plane}

We consider a NC plane given by Hermitian coordinate 
operators $\hat x_{\mu}$, which obey
\begin{equation}
[ \hat x_{\mu}, \hat x_{\nu} ] = i \theta \epsilon_{\mu \nu} \ .
\end{equation}
We impose a (fuzzy) lattice structure on this plane by
means of the operator identity
\begin{equation}
\exp \Big( i \frac{2\pi}{a} \hat x_{\mu} \Big) = \hat 1 \!\! 1 \ .
\end{equation}
Now the periodicity of the momentum components $k_{\mu}$
implies 
\begin{equation}
\theta k_{\mu} / 2a \in \Z \ .
\end{equation}
Hence the lattice is automatically periodic, say over the lattice
volume $N \times N$. Then the non-commutativity parameter corresponds to
\begin{equation}
\theta = N a^{2} / \pi \ .
\end{equation}
We see that the limits to the continuum ($a \to 0$) and to the infinite volume
($Na \to \infty$) should be taken simultaneously, if we want to keep $\theta$
finite. In particular, we are interested in the double scaling limit
$a \to 0$, $N \to \infty$, which keeps $\theta = const.$

\section{The 3d NC $\lambda \phi^{4}$ Model}

In the star product formulation, the action of the 
NC $\lambda \phi^{4}$ model in Euclidean space takes the form
\begin{equation}  % \label{starpro}
S[ \phi ] = \int d^{d}x \, \Big[ \frac{1}{2} \partial_{\mu} \phi
\partial_{\mu} \phi + \frac{m^{2}}{2} \phi^{2} + \frac{\lambda}{4}
\phi \star \phi \star \phi \star \phi \Big] \ . \nonumber
\end{equation}
Since the star product does not affect the bilinear terms,
$\lambda$ determines the strength of NC effects.
We consider this model in $d=3$
with a commutative Euclidean time and a NC plane. Its formulation
on a $N^{3}$ lattice can be mapped onto a matrix model\cite{AMNS} 
of the form
\begin{eqnarray}
&& \hspace*{-7mm} S [ \bar \phi ] = N {\rm Tr} \sum_{t=1}^{N} \left\{ \frac{1}{2}
\sum_{\mu =1}^{2} \Big[ \Gamma_{\mu} \bar \phi (t) \Gamma_{\mu}^{\dagger}
- \bar \phi (t) \Big]^{2} \right. \nonumber \\
&& \hspace*{-7mm} \left. + \frac{1}{2} \Big[ \bar \phi (t+1) - \bar \phi (t) 
\Big]^{2} + \frac{m^{2}}{2} \bar \phi^{2}(t) 
+ \frac{\lambda}{4}\bar \phi^{4}(t) \right\} \ ,  \label{matrix}
\end{eqnarray}
where each time site $t = 1 \dots N $ accommodates a Hermitian
$N \times N$ matrix $\bar \phi (t)$. The ``twist eaters'' $\Gamma_{\mu}$
provide a shift by one lattice unit in a spatial direction, if they
obey the 't Hooft-Weyl algebra
\begin{equation}
\Gamma_{\mu} \Gamma_{\nu} = Z_{\nu \mu} \Gamma_{\nu} \Gamma_{\mu} \ .
\end{equation}
The twist $Z_{\mu \nu} = Z _{\nu \mu}^{*} $ is a phase factor; 
in our formulation it reads $Z_{12} = \exp(i \pi (N+1)/N)$.

\section{The Phase Diagram}

In contrast to the star product formulation, the matrix 
formulation of eq.\ (\ref{matrix}) is suitable for Monte Carlo 
simulations. Our numerical results are described in Ref.\
[\refcite{mainpap}], and in several proceeding contributions
as well as a Ph.D. thesis.\cite{Procs} Similar techniques were
applied to arrive at non-perturbative results for 2d NC
field theories, in particular for the $\lambda \phi^{4}$ model
on a NC plane\cite{AC,mainpap} and on a fuzzy sphere,\cite{XM}
and for NC QED$_{2}$.\cite{2dU1}

In the 3d model described above, we first explored the phase diagram.
We found it to be stable for $N \gsim 25$ in the plane spanned
by the axes $N^{2} m^{2}$ and $N^{2} \lambda$, see Figure 1.

%\begin{figure}[h]
\begin{center}
%\rule{2cm}{0.2mm}\hfill \rule{2cm}{0.2mm}
%\vskip 4.6cm
%\rule{2cm}{0.2mm}\hfill \rule{2cm}{0.2mm}
\psfig{figure=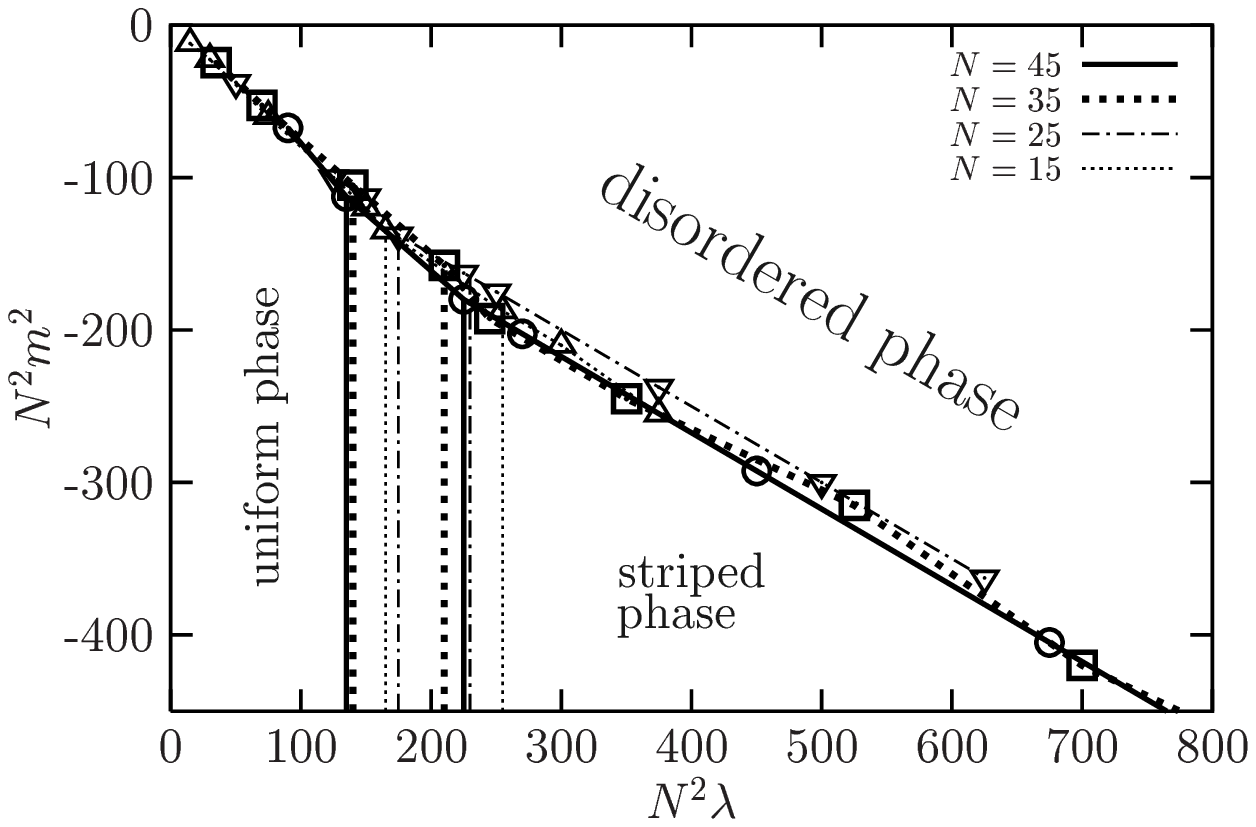,height=2in}
\end{center}
{\bf Figure 1}: {\em The phase diagram of the 3d NC $\lambda \phi^{4}$ model,
obtained from numerical simulations on a $N^{3}$ lattice.}
%\label{fig:radk}
%\end{figure}

\vspace*{2mm}

\noindent
Strongly negative $m^{2}$ leads to ordering. At weak $\lambda$
the order is uniform as in the commutative world, whereas at stronger $\lambda$
(corresponding to an amplified $\theta$) stripe patterns dominate.
This picture agrees with analytic conjectures.\cite{GuSo}
The occurrence of a striped phase is a qualitative difference from the
commutative $\lambda \phi^{4}$ model, though similar effects are
known for instance in the (commutative) Gross-Neveu model at large
chemical potential\cite{GN} and in ferromagnetic superconductors.

Our results for the hysteresis suggest that the uniform-striped 
transition is of first order, while both disorder-order transitions
are of second order.
The formation of stripes was observed by introducing a momentum
dependent order parameter. The problem in this direct
consideration is that at our values of $N$ usually just two stripes 
can be observed as manifestly stable. However, multi-stripe patterns are 
supposed to dominate the striped phase at large volumes.
We will demonstrate this behavior in an indirect way in Section 5.

\section{Dispersion Relation}

The correlation functions with a spatial separation have a fast
but non-exponential decay;\cite{mainpap} apparently it is distorted 
by the NC geometry. In Euclidean time direction, however, the
decay of the correlator turned out to be exponential.
At fixed spatial momenta $\vec p = (p_{1},p_{2})$ this
allows us to determine the energy $E$ and thus
the dispersion relation $E^{2}(\vec p^{\, 2})$.
It is most instructive to look at it in the disordered phase
--- which does not suffer much from finite size effects --- close
to the ordering transition. Figure 2 shows examples close to the uniform
order (on top) and close to the striped order (below). The former
follows the familiar linear shape, whereas the latter has its energy
minimum at non-zero momentum. Clearly for decreasing $m^{2}$ the minimal
mode condenses, giving rise to a stripe pattern.

\begin{center}
\psfig{figure=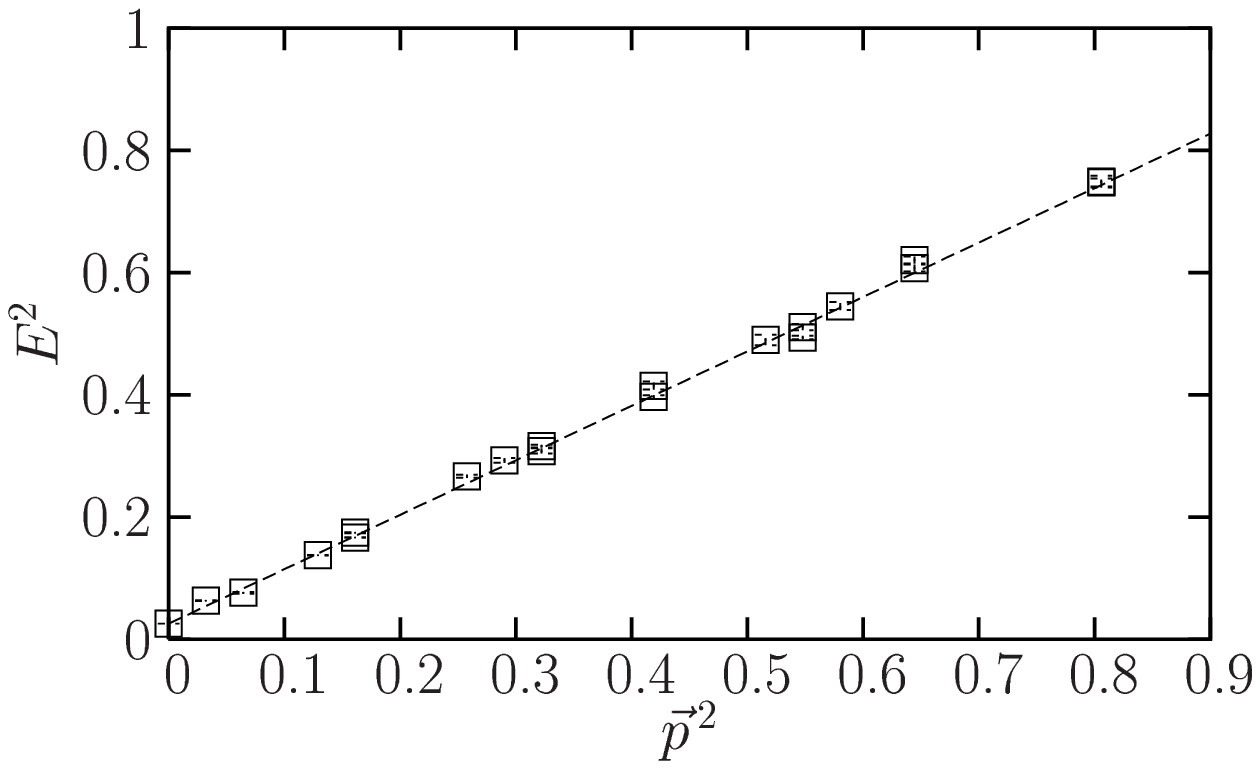,height=1.5in}
\hspace*{1mm} \psfig{figure=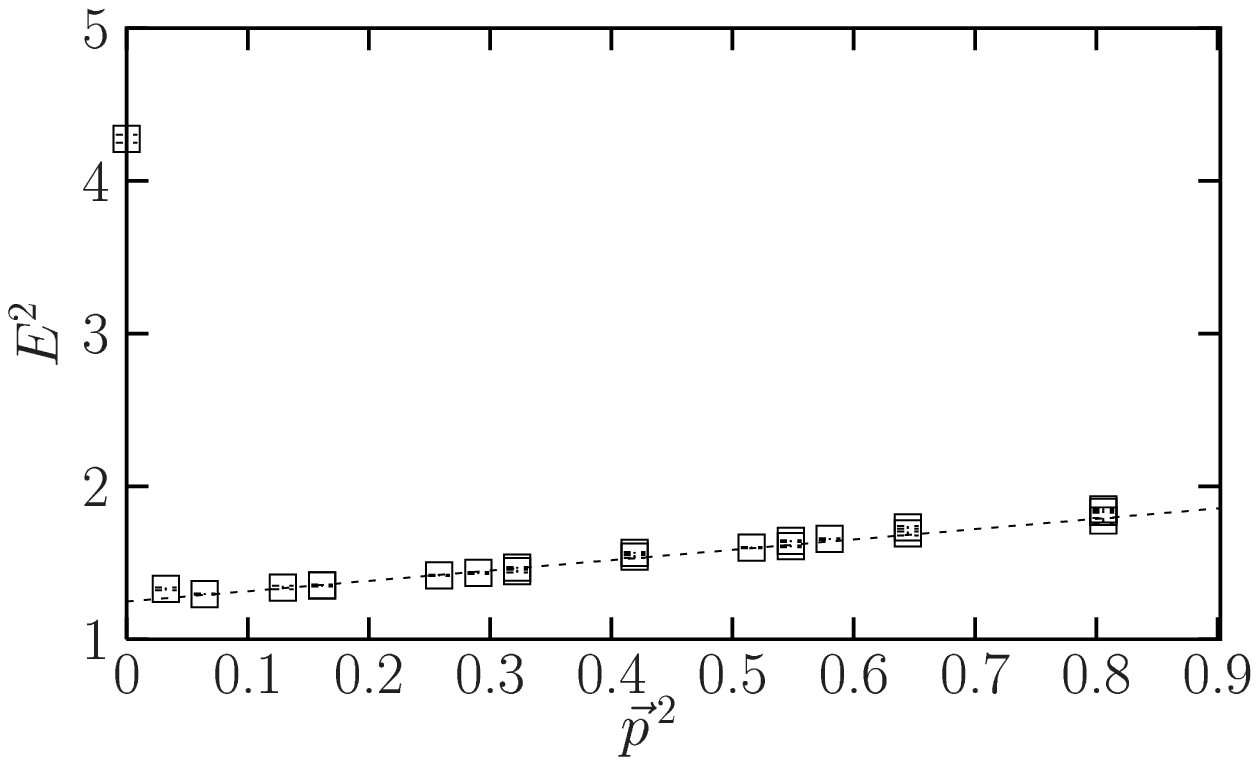,height=1.46in}
\end{center}
{\bf Figure 2}: {\em The dispersion relation in the disordered phase,
close to ordering, at $N=35$. For $\lambda = 0.06$ (on top)
we obtain the usual linear dispersion, but for $\lambda = 100$
(below) the energy minimum moves to a finite momentum.}

\vspace*{-1mm}
\section{Continuum limit}
\vspace*{-1mm}

In order to obtain a ``physical scale'' we have to identify a dimensionful
quantity. To this end we consider the planar limit $N\to \infty$ at
fixed $\lambda$ and $m^{2}$. This pushes the IR jump in the dispersion
towards zero, and we obtain (for all finite momenta) a linear dispersion
of the form $ E^{2} = M_{\rm eff}^{2} + \vec p^{\, 2}$.
Measuring now $M_{\rm eff}^{2}$ at fixed $\lambda$ but varying $m^{2}$
we observed a linear dependence,
\begin{equation}
M_{\rm eff}^{2}\vert_{\lambda \, = \, const.} = \mu^{2} + \gamma m^{2}
\ ,
\end{equation}
which corresponds to the critical exponent $\nu = 1/2$. For instance,
at $\lambda = 50$ we found the critical parameter $m_{c}^{2}
= - \mu^{2}/\gamma = -15.01(8)$.

The continuum limit is now taken such that the dimensionful
effective mass, $M_{\rm eff}/a$, remains constant ($a$ being the lattice
spacing). Hence the double scaling limit means $N \to \infty$ and
$m^{2} \to m_{c}^{2}$ such that $N(m^{2}-m_{c}^{2}) = const.$
In our study we chose the latter constant as $100$, which implies
$\theta = 100 \gamma /\pi = 9.77(6) (a/M_{\rm eff})^{2}$.

It turns out that the dispersion relation stabilizes in this
double limit at all finite momenta, which demonstrates the non-perturbative 
renormalizability of this model. Figure 3 (on top) shows the energy
minimum around $ \vec p^{\, 2}/a^{2} \lsim 0.1$. This translates
into a dominant, finite stripe width, so after condensation we
expect an infinite number of stripes in the double scaling limit
to the continuum and infinite volume.

%\vspace*{-1mm}
\begin{center}
\psfig{figure=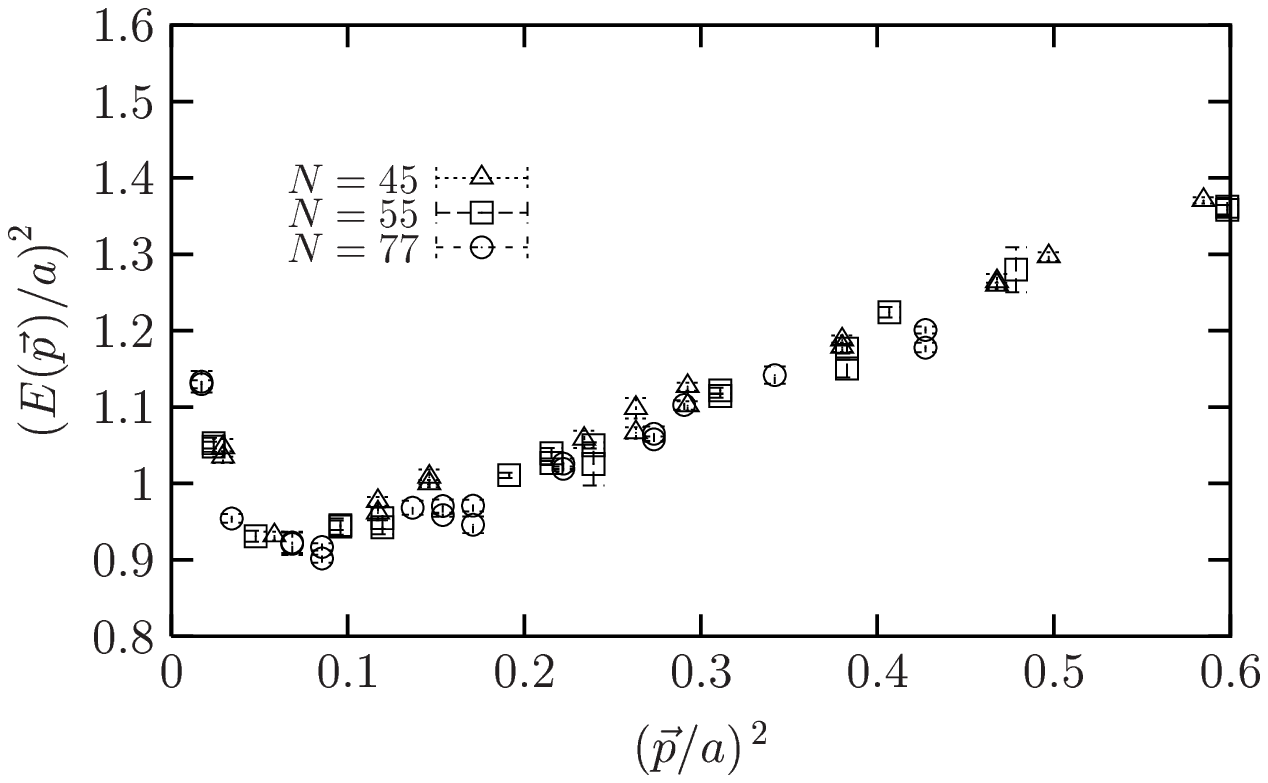,height=1.5in}
\hspace*{1mm} \psfig{figure=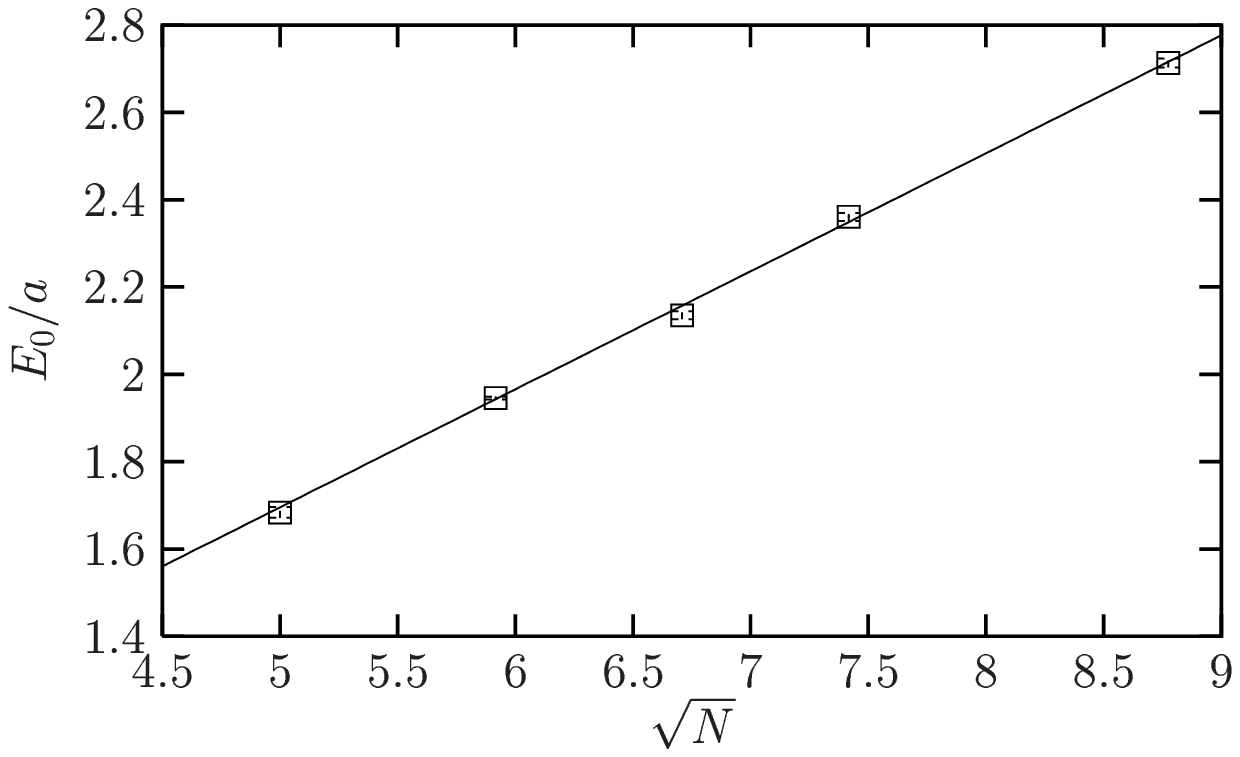,height=1.46in}
\end{center}
\vspace*{-1mm}
{\bf Figure 3}: {\em On top: The dispersion relation as we approach the 
double scaling limit. 
%to the continuum and infinite volume, at
%constant $\theta$. 
Its stabilization indicates non-perturbative
renormalizability. %In particular the stable dip
The dip at finite $(\vec p /a)^{2}$
shows that the striped phase survives this limit.
Below: $E_{0}/a$ diverges linearly in 
$\sqrt{N} \propto 1/a$, in agreement with UV/IR mixing.}

\vspace*{2mm}

\noindent
In this limit, the rest energy $E_{0}/a = E (\vec p = \vec 0 )/a$ diverges
linearly in $\sqrt{N} \propto 1/a$, see Figure 3 (below).
Also the UV divergence is linear in this model, hence our
observation agrees perfectly with the concept of UV/IR mixing,
which is known from perturbation theory.\cite{UVIR} Here this mixing
is observed as a non-perturbative effect, so it belongs to the
very nature of the system.

\vspace{-1mm}
\section{Conclusions}
\vspace{-1mm}

We have shown that the NC 3d $\lambda \phi^{4}$ model --- with two NC 
spatial direction and a Euclidean time --- is {\em non-perturbatively 
renormalizable}, and that its phase diagram includes a {\em striped
phase}, which is there to stay in the continuum limit (more precisely:
in the double scaling limit to zero lattice spacing and infinite volume,
at a fixed non-commutativity parameter $\theta$).

The striped phase implies the spontaneous breaking of translation symmetry.
It also exists in the 2d version of this model (omitting the
time direction);\cite{AC,mainpap} note that NC field theories are
non-local, hence the Mermin-Wagner Theorem does not apply.
At $\theta \to \infty$ perturbation theory suggests a commutative
behavior (in this case the equivalence to a large $N$ matrix field theory). 
However, in our phase diagram the uniform order does not
return at large $\lambda$ (which corresponds to large $\theta$);
in the case of spontaneous symmetry breaking the perturbative
argument is not valid, because it does not capture an expansion
around the striped ground state.\cite{largetheta} \\

\vspace{-1mm}
%\section*{Acknowledgment}
{\it W.B.\ would like to thank the organizers of the 11th Regional Conference on
Mathematical Physics in Tehran for their kind hospitality.}

\end{multicols}

\end{document}